\documentclass[final,5p,times,preprint]{elsarticle}
\usepackage[utf8]{inputenc}
\usepackage{amsfonts}
\usepackage{amsmath}
\usepackage{amssymb}
\usepackage{bm}
\usepackage{lineno,hyperref}
\usepackage{comment}
\usepackage{xcolor}
\usepackage{booktabs}
\usepackage{multirow}
\usepackage[normalem]{ulem}

\newcommand{\vecrp}{\vec{r}\mkern2mu\vphantom{r}'}

\newcommand{\vecKp}{\vec{K}\mkern2mu\vphantom{K}'}

\newcommand{\vecRp}{\vec{R}\mkern2mu\vphantom{R}'}

\journal{Physics Letters B}

\biboptions{comma,sort&compress}

\begin{document}

\begin{frontmatter}

\title{Investigating the $^{10}$Li continuum through $^{9}$Li$(d,p)^{10}$Li reactions}


\author[FAMN]{A. M. Moro\corref{mail}}
\cortext[mail]{Corresponding author}
\ead{moro@us.es}

\author[Padova]{J. Casal}

\author[FAMN]{M. G\'omez-Ramos}

\address[FAMN]{Departamento de F\'{\i}sica At\'omica, Molecular y Nuclear, Facultad de F\'{\i}sica, Universidad de Sevilla, Apartado 1065, E-41080 Sevilla, Spain}


\address[Padova]{Dipartimento di Fisica e Astronomia ``G. Galilei'' e INFN - Sezione di Padova, Via Marzolo 8, I-35131 Padova, Italy
}

\begin{abstract}
The continuum  structure of the unbound system $^{10}$Li, inferred from the  $^{9}$Li$(d,p)^{10}$Li transfer reaction, is reexamined.  Experimental data for this reaction, measured at two different energies, are analyzed with the same reaction framework and structure models.  It is shown that the seemingly different features observed in the measured excitation energy spectra can be understood as due to the different incident energy and angular range covered by the two experiments.  
The present results support the persistence of the $N=7$ parity inversion  beyond the neutron dripline as well as the splitting of the well-known low-lying $p$-wave resonance. Furthermore, they provide indirect evidence that most of the $\ell=2$ single-particle strength, including possible $d_{5/2}$ resonances, lies at relatively high excitations energies.    
\end{abstract}


\end{frontmatter}


\section{Introduction}

Understanding the nuclear shell evolution as a function of the proton-neutron asymmetry is one of the major goals in nowadays nuclear physics. Within this broad and ambitious program, the $N=7$ isotopic chain has received much attention both experimentally \cite{Kry93,You94,Nak94,Thoe99,Cag99,Che01,San03,Pal03,Fuk04,Rog04,Jep06,Sim07,Sum07,Aks08,AlK13,Kwa14,San16,Gur16,Ube16,Cav17,Pes17} and theoretically \cite{Kat99,Tim99,Kan02,Bark04,Blan07,Orr09,For16,Bar17,Vor18}. The  $^{10}$Li system represents  a prominent member of this chain, due to its peculiar features. First, it is the first unbound $N=7$ isotone, following the weakly-bound $^{11}$Be nucleus. 
Second, several experiments \cite{Zin95,Thoe99,Jep06} suggest that its ground state consists of an $\ell=0$ virtual state, followed by a narrow $p$-wave resonance, whose energy sequence would  point toward a persistence of the parity inversion observed in $^{11}$Be. Finally, an accurate knowledge of the  $^{10}$Li system is crucial for a proper understanding of the $^{11}$Li nucleus, the archetypal three-body Borromean nucleus.    


Despite this interest, and the extensive experimental and theoretical efforts, important questions  regarding the structure of $^{10}$Li remain unanswered.  Due to the non-zero spin of the $^{9}$Li {\it core}, the $s$-wave and $p$-wave structures are expected to split into   $(1^-,2^-)$ and  $(1^+,2^+)$ doublets, respectively. However, these doublets have not yet been clearly identified experimentally. In particular, it is unclear whether the prominent peak observed in several experiments~\cite{Sim07,Aks08,Cav17}, and identified with the  $p_{1/2}$ resonance, corresponds to the centroid of the (unresolved) doublet or just to one of its members, with the other component being pushed at higher excitation energies. 

In the case of the $s_{1/2}$ virtual state, the situation is less clear. Experimentally, its presence was inferred from the narrow width of the momentum distribution in one-proton and one-neutron removal  experiments of energetic $^{11}$Be and $^{11}$Li beams  on a carbon target~\cite{Zin95}. Another experimental evidence came from the measurement of the relative velocity distribution between the $^{9}$Li and the neutron resulting from the decay of $^{10}$Li produced after the collision of  a $^{18}$O beam on a $^{9}$Be target~\cite{Thoe99}. This relative velocity was found to peak at zero, which is consistent with an $\ell=0$ configuration for the $^{10}$Li ground-state. The search for this virtual state has been also pursued with transfer experiments. For example, the excitation function extracted for the reaction  $^{9}$Li($d$,$p$)$^{10}$Li measured at $E=2.4$~MeV/u at REX-ISOLDE exhibited an excess of strength at zero energy which was consistent with a virtual state with a (negative) scattering length of the order of 13-24~fm \cite{Jep06}.  
However, a more recent experiment for the same transfer reaction performed at TRIUMF at a higher incident energy \cite{Cav17} did not show any indication of such near-threshold structure, putting into question its very existence. 

The situation regarding the presence of one or more $d_{5/2}$ low-lying resonances is even more controversial. Evidence of such a resonance at $E_r \sim 1.5$~MeV has been reported in a fragmentation experiment of $^{11}\text{Li}$ on $^{12}\text{C}$ performed at GSI \cite{Sim07}, and supported by the theoretical analysis of Blanchon {\it et al.}~\cite{Blan07}. The excitation function extracted from the  $^{9}$Li($d$,$p$)$^{10}$Li reaction \cite{Cav17} displayed also a small bump at $E_r=1.5$~MeV, but the theoretical analysis performed in that work pointed towards an $s$-wave dominance. However, a second bump at $E_r=2.9$~MeV was observed in this experiment with significant $d_{5/2}$ content. 

In the present work, we reexamine the problem of the $^{10}$Li continuum, by presenting a joint and consistent analysis of the  $^{9}$Li$(d,p)^{10}$Li reactions measured at REX-ISOLDE \cite{Jep06} and TRIUMF \cite{Cav17}, using in both cases the same reaction framework and structure models. From this analysis, we conclude that both measurements can be described consistently using the same model for the $n$-$^9$Li interaction, which has also been able to successfully reproduce experimental data for $^{11}$Li$(p,d)^{10}$Li \cite{Cas17} and $^{11}$Li$(p,pn)^{10}$Li \cite{Gom17}.

This work is organized as follows. In Sec.~\ref{sec:calc} the details of the reaction framework and the structure models used are presented. In Sec.~\ref{sec:results}, the results of the calculations are shown and compared to experimental data, focusing on the compatibility of the data from \cite{Jep06} and \cite{Cav17} and on the possible existence of a low-energy $d$-wave resonance. Finally, in Sec.~\ref{sec:sum} the summary and conclusions of this work are detailed.

\section{Details of the calculations}\label{sec:calc}
\subsection{Reaction framework}
To describe the  $^{9}$Li($d$,$p$)$^{10}$Li reaction, we employ the Transfer to the Continuum (TC) formalism \cite{Mor06,Jep06}, which is based on the prior-form transition amplitude for unbound final states.  Denoting this reaction as $A(d,p)B$, this transition amplitude is expressed as:
\begin{equation}
\mathcal{T}_{if}= \langle \Psi_{f}^{(-)}(\vecRp,\vecrp)|V_\text{nA}+U_\text{pA}-U_\text{dA} | \phi_d(\vec{r}) \chi_{dA}^{(+)}(\vec{R}) \rangle,
\label{eq:tmatrix}
\end{equation}
where $A$ and $B$ denote in our case the $^{9}\text{Li}$ and $^{10}\text{Li}$ systems, $U_\text{pA}$ and $U_\text{dA}$ are optical potentials for the $p$-$A$ ad $d$-$A$ systems, respectively and $V_\text{nA}$ is the interaction describing the $^{10}$Li continuum. The function $\chi^{(+)}_{dA}$ is the distorted wave generated by the optical potential $U_{dA}$ and $\Psi^{(-)}_f(\vecRp,\vecrp)$ is the exact three-body wave function for the outgoing $p$+$n$+$^{9}$Li three-body system, with $\vecrp$ and $\vecRp$ denoting the $n$-$^{9}$Li and $p$-$^{10}$Li relative coordinates, respectively.  The $\pm$ superscripts refer to the usual incoming or outgoing boundary conditions. The wave function $\Psi_{f}^{(-)}(\vecRp,\vecrp)$ is the time-reversed of $\Psi_{f}^{(+)}(\vecRp,\vecrp)$, which satisfies the three-body equation:
\begin{equation}
\label{eq:sch}
[\hat{T}_{\vecRp} + \hat{T}_{\vecrp} + V_\text{pn} + U_\text{pA} + V_\text{nA} -E]    \Psi_{f}^{(+)}(\vecRp,\vecrp)=0 ,
 \end{equation}
 with $\hat{T}_{\vecRp}$ and $\hat{T}_{\vecrp}$ representing the kinetic energy operators associated, respectively, to the $p+^{10}$Li and $n+^{9}$Li relative motion, and $E$ the total energy of the system. To solve the latter equation, the wavefunction $\Psi^{(-)}_f(\vecRp,\vecrp)$ is expanded 
in  $n+^9$Li states with well-defined energy and angular momentum, as in the continuum-discretized coupled-channels (CDCC) method \cite{Aus87}. Therefore, for each total angular momentum and parity of the $^{10}$Li system ($J^\pi$), the $n+^{9}$Li continuum is discretized in energy {\it bins}. Denoting each of these  $^{10}$Li discretized states by $\psi^{i,J,M}_B(\xi_A, \vecrp) = [ \phi_I(\xi_A)  \otimes \varphi_{i,\ell,j}(\vecrp) ]_{J M}$, where $\phi_I(\xi_A)$ is the core ($^9$Li) internal wavefunction with intrinsic spin $I$ and $\varphi_{i,\ell,j}(\vecrp)$ the neutron-$^9$Li relative wavefunction, the CDCC-expansion of the final wavefunction can be written as
 \begin{equation}
 \label{eq:PhiCDCC}
      \Psi_{f}^{(-)}(\vecRp_i,\vecrp) \simeq \sum_{i,J^\pi} \chi_{i,J^\pi} (\vecKp_i,\vecRp) \psi^{i,J,M}_B(\xi_A, \vecrp),
 \end{equation}
 where  $\vecKp_i$ is the final momentum of the outgoing proton in the CM frame for a particular final (discretized) state $i$ of the $n+^{9}$Li system.  Inserting the expansion (\ref{eq:PhiCDCC}) into the Schr\"odinger equation (\ref{eq:sch}) gives rise to a set of coupled differential equations for the unknown functions $\chi_{i,J^\pi} (\vecKp_i,\vecRp)$.  These calculations were performed with the code FRESCO \cite{Thom88}.

\subsection{$^{10}$Li structure models}
A key input of the present calculations is the $^{10}$Li structure model which, within the two-body model assumed here, is specified by the $n+^{9}$Li interaction. In the original analysis of $^{9}$Li$(d,p)^{10}$Li data performed in \cite{Jep06}, a simple Gaussian interaction was adopted, with central and spin-orbit terms, whose strengths were adjusted to give a near-threshold virtual state ($\ell=0$) and a $p$-wave resonance at $E_r \sim 0.4$~MeV. The spin or the $^{9}$Li was ignored and higher waves were not considered. Although this simple model provided a satisfactory description of the data from that work, in our recent analysis of the  $^{11}$Li$(p,pn)^{10}$Li reaction~\cite{Cas17} it was shown that a proper description of the $^{10}$Li excitation energy profile required the inclusion of a more realistic model, including the $^{9}$Li spin which, due to the spin-dependent interaction with the valence neutron, gives rise to a splitting of the $s$-wave virtual state and $p$-wave resonance into $1^-/2^-$ and $1^+/2^+$ doublets, respectively. The adopted model, referred to as P1I, was also found to provide a robust description of the $^{11}$Li$(p,d)^{10}$Li transfer reaction~\cite{Gom17}. Here, we will consider the same model to describe, simultaneously, the $^{9}$Li$(d,p)^{10}$Li experimental data from Refs.~\cite{Jep06,Cav17}. Our parametrization, based the shallow potential in Ref.~\cite{Gar03}, consists of $\ell$-dependent central, spin-orbit and spin-spin components,
\begin{equation}
\label{eq:P1I}
V_{nA}^\ell (r) = V_c^\ell(r)+V_{\text{so-v}}^\ell(r)\vec{\ell}\cdot\vec{s}_n+V_{\text{so-c}}^\ell(r)\vec{\ell}\cdot\vec{I} + V_{\text{ss}}^\ell(r)\vec{s}_n\cdot\vec{I},
\end{equation}
where $\vec{s}_n$ and $\vec{I}$ are the intrinsic spins of the valence neutron and the $^9$Li core, respectively. We use Gaussian shapes for the radial functions, $V_i^\ell(r)=v_i^\ell\exp{[-\left(r/R\right)^2]}$, with $R$ fixed to 2.55~fm. As discussed in Ref.~\cite{Cas17}, the depths $v_c^0$ and $v_{\text{ss}}^0$ are chosen so that only the $2^-$ state in the $s$-wave doublet exhibits a marked virtual-state character. For the $p$ waves, $v_{\text{so-v}}^1$ is fixed to a large positive value to suppress the $p_{3/2}$ Pauli forbidden states. Then, the depth of the central potential $v_c^1$ is adjusted to give the $1^+/2^+$ centroid around 0.5 MeV, and $v_{\text{ss}}^1,v_{\text{so-c}}^1$ are used to produce a small $p-$wave splitting with the $1^+$ resonance at lower energy than the $2^+$. With this choice, our level sequence is analogous to that in Ref.~\cite{Kat99}. Our final set of parameters is: $v_c^{0,1}=-5.4,260.75$ MeV, $v_{\text{ss}}^{0,1}=-4.5,1.0$ MeV, $v_{\text{so-v}}^1=300.0$ MeV and $v_{\text{so-c}}^1=1.0$ MeV. Note that the P1I model, as originally devised in Refs.~\cite{Cas17,Gom17}, has no $\ell=2$ term. We will discuss about $d$-waves in Sec.~\ref{sec:dwave}.

In order to illustrate the importance of the splitting due to the spin of $^9$Li, we consider also a  model which does not consider explicitly the $^{9}$Li spin ($I=0$) and includes only central and spin-orbit terms, i.e.,
 \begin{equation}
\label{eq:P3}
 V_{nA}^\ell (r) = v_c^\ell f(r,R,a)+ v_{\text{so-v}}^\ell \frac{1}{r}\frac{df(r,R,a)}{dr}\vec{\ell}\cdot\vec{s}_n ,
 \end{equation}
where the depths are given by $v_c^{0,1}=-50.5$, $-39.0$ MeV, $v_{\text{so-v}}^1=40.0$ MeV~fm$^2$ and the radial factors $f(r,R,a)$ are of Woods-Saxon type and use the parameters  $R=2.642$ fm and $a=0.67$ fm. This corresponds to the $\ell=0,1$ part of the potential referred to as P3 in Refs.~\cite{Cas17,Gom17}.
 
\section{Numerical results}\label{sec:results}
\begin{figure}[!hbt]
\centering
\includegraphics[width=0.75\linewidth]{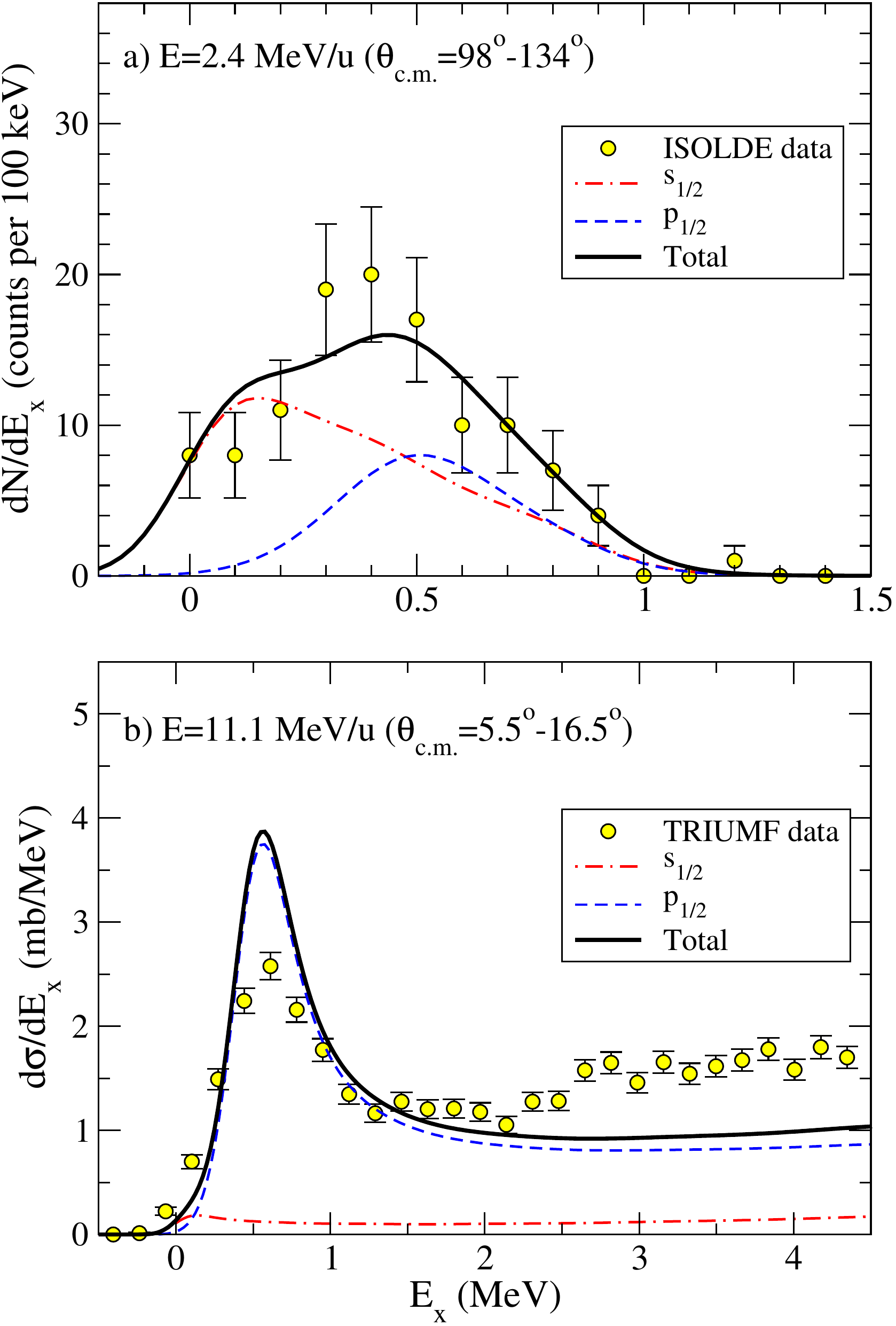}
\caption{\label{fig:dsde_p3} (Color online) Excitation energy spectrum for  $^{9}$Li(d,p)$^{10}$Li at  $E=2.36$~MeV/u (top) and  $E=11.1$~MeV/u (bottom). The data from Refs.~\cite{Jep06,Cav17} are compared with TC calculations using the structure model P3 (see text), including only $s$ and $p$ waves. Calculations are folded with the experimental energy resolution for each reaction. }
\end{figure}

We present first calculations using the $^{10}$Li model without $^{9}$Li spin using the parameters of the potential P3 introduced in the previous section. In Fig.~\ref{fig:dsde_p3} we show the excitation energy functions of $^{10}$Li from the  $^{9}$Li$(d,p)^{10}$Li experiments of Refs.~\cite{Jep06} and \cite{Cav17} at  $E=2.36$~MeV/u (top) and  $E=11.1$~MeV/u (bottom), respectively.  The data are compared with reaction calculations using the TC formalism outlined in the previous section. For the $E=2.36$~MeV/u case, we adopt the  same deuteron and proton optical potentials used in the calculations of  Ref.~\cite{Jep06}. For the $E=11.1$~MeV/u case, we employ the deuteron potential from the Perey and Perey compilation \cite{Per76} and the proton optical potential from the global parametrization of Koning and Delaroche \cite{KD03}. Note that the calculations are integrated over the angular range covered by each experiment (indicated in the labels), and convoluted with the corresponding experimental resolution, using Gaussians with FWHM=250~keV and FWHM=200~keV for the REX-ISOLDE and TRIUMF data, respectively. In the lower energy case, the excitation function is multiplied by the acceptance function~\cite{Jep06}, which converts the calculated cross section into counts per energy interval. This acceptance falls quickly to zero as the excitation energy increases (i.e., decreasing proton energy) vanishing for excitation energies above $\sim$1~MeV and thus has an important impact on the information that can be extracted from these data. In particular, any hypothetical $d$-wave resonance,  predicted theoretically~\cite{Nun96,Kat99}  and suggested  experimentally \cite{Sim07}, would not be visible in these data. 

In each panel of  Fig.~\ref{fig:dsde_p3} we show the separate $\ell=0$ ($s_{1/2})$ and $\ell=1$ ($p_{1/2}$) contributions, as well as their sum. In the case of the REX-ISOLDE data, it is seen that the calculations reproduce rather well the excitation function. The inclusion of the $\ell=0$ virtual state is essential to account for the excess of strength near zero energy, whereas the $\ell=1$ resonance contributes mainly to the peak at $E_x\sim0.4$~MeV. These results are in accord with the conclusions of Ref.~\cite{Jep06}.

Regarding the TRIUMF data (Fig.~\ref{fig:dsde_p3}b),  the excitation energy function is dominated by prominent peak at $E_x\sim0.45$~MeV followed by a flatter contribution extending to higher excitation energies. Our calculations indicate that the peak is almost exclusively due to the $\ell=1$ resonance, whereas the virtual state gives an almost negligible contribution in the whole energy range. These results are in agreement with those reported in the original analysis of Ref.~\cite{Cav17}.  For excitation energies above $\sim1$~MeV the calculations underestimate the data, which might might be due to the contribution of higher ($\ell>1$) waves of the $^{10}$Li continuum. This will be considered in the following section.

\begin{figure}[!hbt]
\centering
\includegraphics[width=0.75\linewidth]{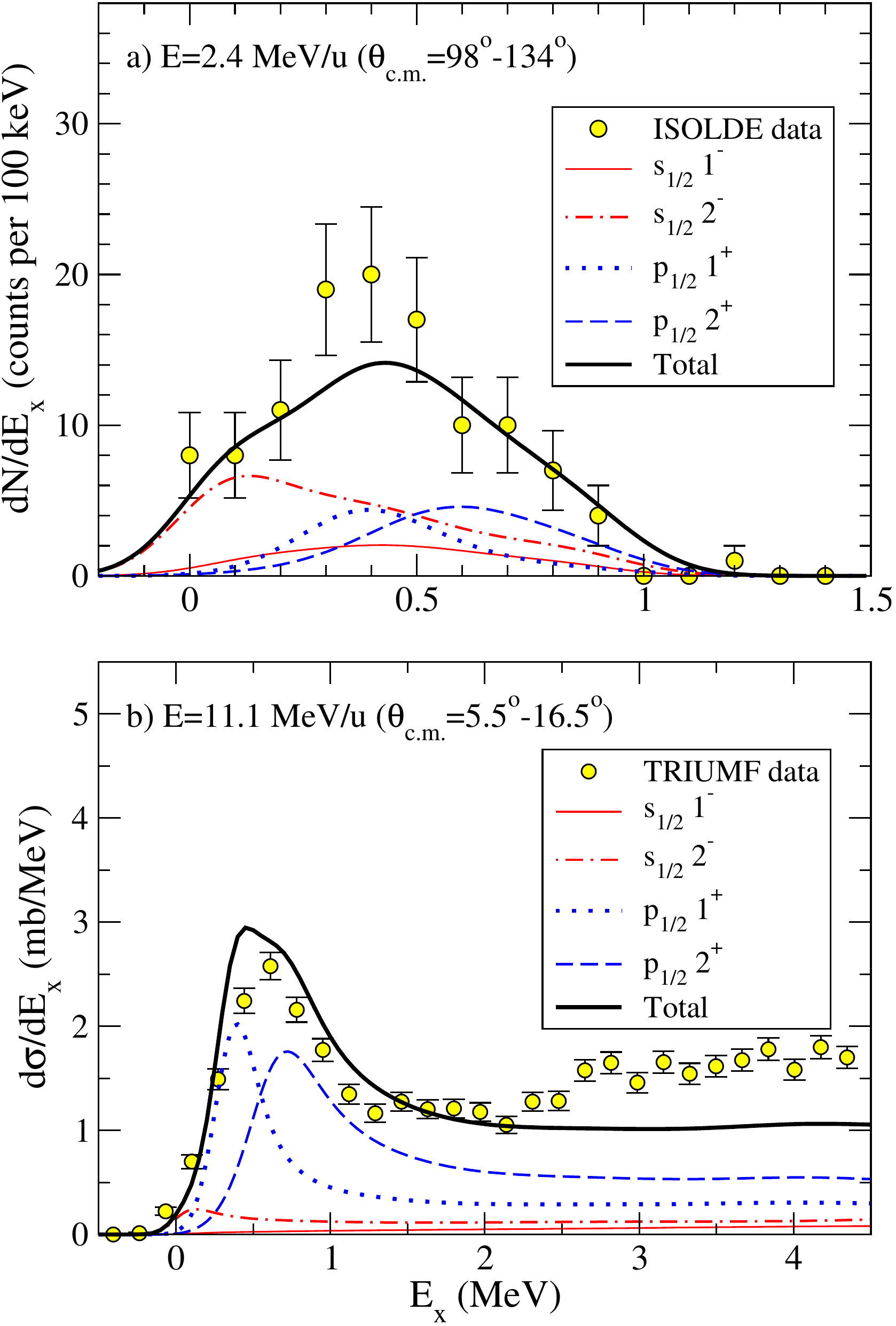}
\caption{\label{fig:dsde_p1I} (Color online) Same as Fig.~\ref{fig:dsde_p1I}, but using the $^{10}$Li model P1I (see text), which includes the spin of the $^{9}$Li core.}
\end{figure}

Although the calculations shown in Fig.~\ref{fig:dsde_p3}b provide a reasonable account of the resonant peak observed in the TRIUMF data, the predicted energy profile is found to overestimate the height of this peak and exhibits a narrower shape. 
A possible reason for this discrepancy is the omission of the $^{9}$Li spin in the P3  model which, as noted before, will naturally lead to a fragmentation of the single-particle resonances into spin multiplets. 
To examine the influence of this fragmentation on the data, we have repeated the calculations using the P1I model given by Eq.~(\ref{eq:P1I}), which accounts for this  splitting in a phenomenological way.
In Fig.~\ref{fig:dsde_p1I}, the TC calculations based on this  $^{10}$Li  model are compared with the data from the two considered experiments. In the case of the REX-ISOLDE data, the results are  very similar to those obtained with the P3 model, indicating that these data are not sensitive to this single-particle fragmentation. This is partly due to the limited energy resolution and the low statistics of these data. The situation is different for the TRIUMF data. In this case, the splitting of the $p_{1/2}$ resonance results in a broadening of the resonant peak and a reduction of its magnitude, improving the agreement with the data. Note that the contribution of the $\ell=0$ wave remains negligible with the new potential.

\begin{figure}[!hbt]
\centering
\includegraphics[width=0.75\linewidth]{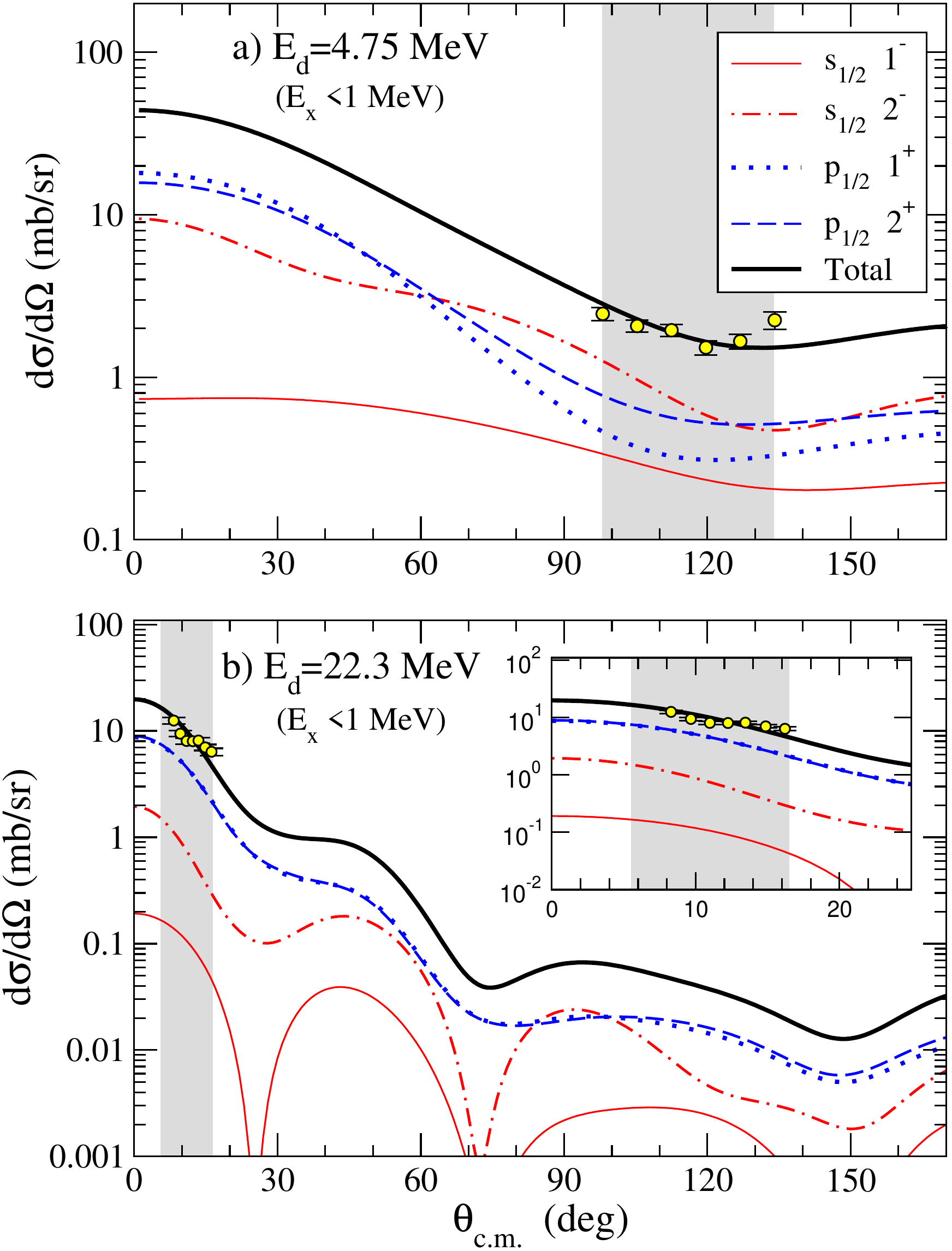}
\caption{\label{fig:dsdw} (Color online) Differential angular distributions for  $^{9}$Li$(d,p)^{10}$Li at $E=2.36$~MeV/u (top) and $E=11.1$~MeV/u (bottom) integrated up to an excitation energy of $E_x =1$~MeV.  The shaded areas highlight the angular regions used in the experiments of Refs.~\cite{Jep06} and \cite{Cav17} to extract the excitation energy function. The inset  of the bottom panel shows a zoom on the experimental angular region.}
\end{figure}

The seemingly different role played by the $s$-wave continuum in the two reactions seems to lead to conflicting results regarding the structure of the $^{10}$Li continuum. The difference can be however understood by looking at the corresponding angular distributions. This is illustrated in Fig.~\ref{fig:dsdw}, where the calculated angular distributions (using the P1I model) are plotted for the two energies. The shaded regions correspond to the angular range used for the extraction of the experimental excitation functions. From the upper panel of this figure, it becomes apparent that, for the angular range spanned by the REX-ISOLDE experiment, \cite{Jep06} the $s$- and $p$-wave contributions are of similar magnitude, and hence both contributions are visible in the excitation energy spectrum. Interestingly, for this incident energy the calculations predict a negligible contribution of the $s$-wave for $\theta_\text{c.m.}<60^\circ$.  In the case of the TRIUMF experiment (bottom panel), the  measured angular distribution corresponds to small angles which, according to the present calculations, are largely dominated by the $p$-wave contribution. This result explains why the analysis of Ref.~\cite{Cav17} did not find any evidence of the low-lying $s$-wave strength in these data. 

\subsection{Evidence for $d$-wave resonances}\label{sec:dwave}
The comparison of Fig.~\ref{fig:dsde_p1I} show  a clear underestimation of the data  from Ref.~\cite{Cav17}  for excitation energies above $\sim$2~MeV.  This might be due to additional contributions from the $^{10}$Li continuum, with the $\ell=2$  wave being a natural candidate. 
Guided by previous experimental evidences and theoretical predictions pointing toward the existence of one or more low-lying $\ell=2$ resonances, we have performed additional  calculations at $E=11.1$~MeV/u, including $\ell=0,1,2$ waves. It must be noted, however,  that the P3 and P1I models adopted in this work were developed in Ref.~\cite{Cas17} making use only of $\ell=0,1$ waves and, owing to their phenomenological nature,  do not provide an obvious extension to $\ell>1$ waves.  In this same spirit, we have performed additional TC calculations based on these P3 and P1I models, including also  a $\ell=2$ component   with the same geometry as the $\ell=0,1$ parts, but with the strengths adjusted to produce resonances at several excitation energies.  

\begin{figure}[!h]
\centering
\includegraphics[width=0.75\linewidth]{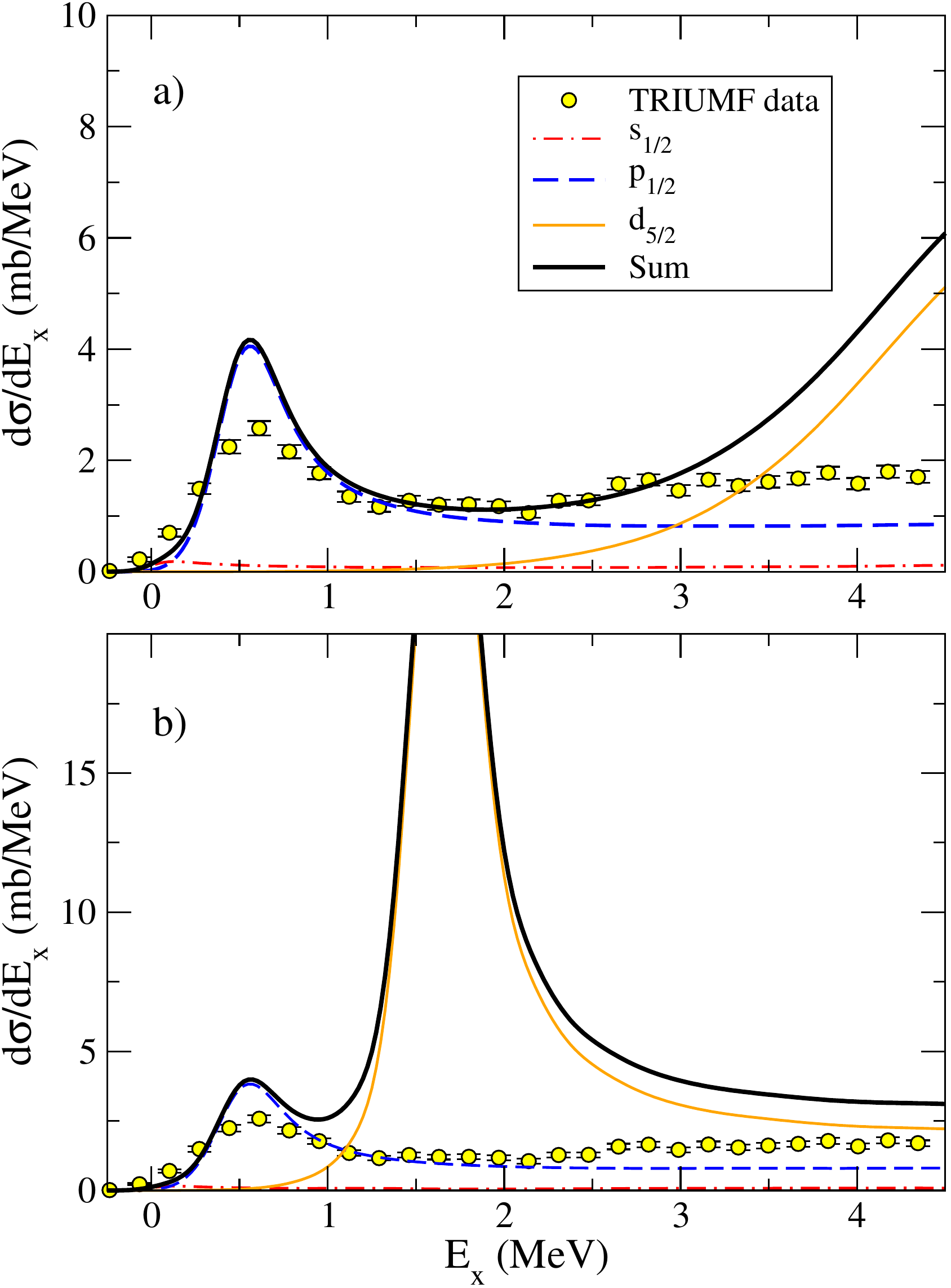}
\caption{\label{fig:dsde_p3_dres} (Color online)  Excitation energy function for the $^{9}$Li$(d,p)^{10}$Li reaction at $E=11.1$~MeV/u. Experimental data from \cite{Cav17} are compared with TC calculations omitting the $^{9}$Li spin, using an extended P3 model with an  hypothetical $d_{5/2}$ resonance at: a) $E_x=4.5$~MeV and b) $E_x=1.5$~MeV. 
}
\end{figure}

\begin{figure}[!ht]
\centering
\includegraphics[width=0.75\linewidth]{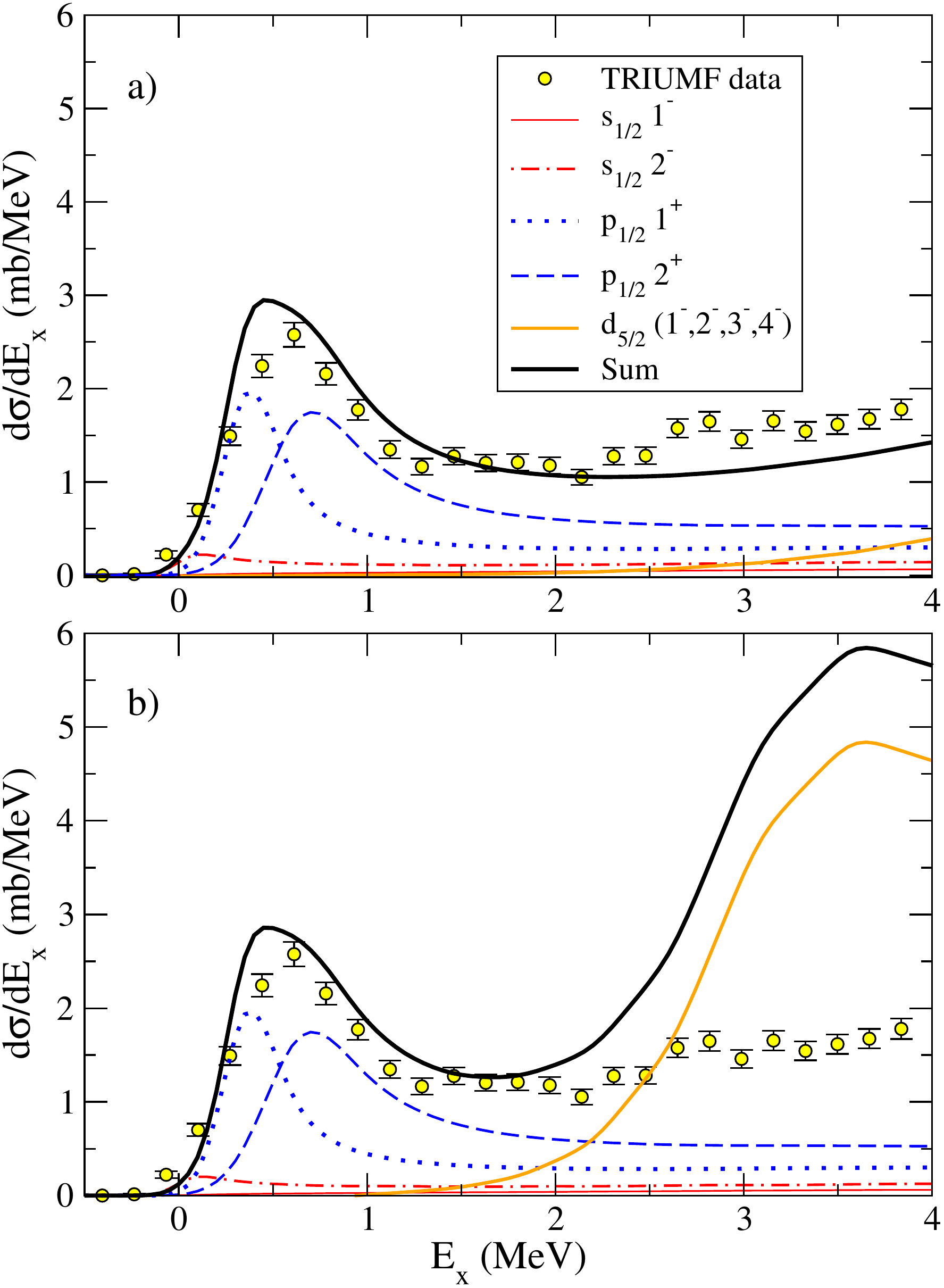}
\caption{\label{fig:dsde_p1I_dres} (Color online)
Same a Fig.~\ref{fig:dsde_p3_dres}, but for an extended P1I model including $d_{5/2}$ waves for the $^{10}$Li continuum. In panel a) the $d$-wave model results in a broad resonance-like structure with a peak at 6~MeV for the $4^-$ component, while panel b) corresponds to a $4^-$ resonance at 3.9~MeV. For clarity, the four contributions ($1^-, 2^-, 3^-, 4^-$) arising from the coupling of the $d_{5/2}$ configuration with the core spin are presented summed.  
}
\end{figure}

The results are shown in Figs.~\ref{fig:dsde_p3_dres} and \ref{fig:dsde_p1I_dres} for the extended P3 and P1I models, respectively.  In Fig.~\ref{fig:dsde_p3_dres}a, the strength of the $d$-wave is adjusted to produce a $d_{5/2}$ resonance at $E_x=4.5$~MeV. Below $E_x=2$~MeV, the $d$-wave does not contribute but, above this energy, it produces a steady increase of the cross section, which is not consistent with the trend of the data. We have  considered also the case of a $d_{5/2}$ resonance at $E_x=1.5$~MeV, following the suggestions of Refs.~\cite{Sim07,Blan07}. In this case, as shown in Fig.~\ref{fig:dsde_p3_dres}b, the excitation function is dominated by a pronounced resonant peak at the nominal energy of the resonance, which is clearly incompatible with the data. It is interesting to note that this model gave a reasonable account of the $^{9}$Li+n decay spectrum from the one-neutron removal data for the reaction $^{11}\text{Li}(p,pn)$ according to the analysis performed in \cite{Gom17} (potential P5 in that reference). The results of the present work strongly suggest that the aforementioned  agreement was probably fortuitous. 

Regarding the calculations based on the extended P1I model, we show the results in  Fig.~\ref{fig:dsde_p1I_dres}. For the $\ell=2$ interaction, we have kept the same geometry as for $\ell=0,1$, varying the depth of the central part for $\ell=2$ in order to produce different spectra for $^{10}$Li. In panel a), the interaction has been adjusted to give a reasonable reproduction of the high energy tail of the experimental data, resulting in a broad resonance-like structure for the $4^-$ wave at 6~MeV, consistent with prediction of Ref.~\cite{Kat99}. In panel b) the interaction has been adjusted to produce a resonance at 3.9~MeV for the $4^-$ wave. As can be seen in the figure, for this latter model, the cross section is severely overestimated in the area of the resonance.

It is worth noting that the type of resonances considered here are of single-particle nature. Consideration of additional degrees of freedom, such as pairing and core excitations,   might give rise to more complicated, many-body resonances \cite{Orr09,Vin95,Nun96,Bro04}.  
Although our calculations do not completely rule out the existence of this type of resonances in the explored $^{10}$Li continuum, they strongly suggest that the majority of the $\ell=2$ strength is concentrated at higher excitation energies.

\section{\label{sec:sum} Summary and conclusions}
In summary, we have performed a joint analysis of the data from two $^9$Li$(d,p)^{10}$Li experiments at $E=2.4$~MeV/u (REX-ISOLDE) and $E=11.1$~MeV/u (TRIUMF), populating the low-lying continuum of $^{10}$Li. To that end, we have employed the transfer-to-the-continuum reaction framework and several $^{10}$Li structure models, finding that both experiments can be described with the same  $^{10}$Li model. These calculations show that, for the lower energy experiment, the $s$-wave virtual state in $^{10}$Li plays a key role, and is essential to explain the near-threshold strength. By contrast, for the higher energy experiment, the excitation energy below $E_x<1$~MeV is largely dominated by the $p_{1/2}$ resonance. The different importance of the $s$-wave virtual state can be explained considering the different angular ranges covered by the two experiments. Furthermore, the comparison of the calculations with the low-energy peak of the TRIUMF data suggests the fragmentation of the $p_{1/2}$ resonance into $1^+$ and $2^+$ components arising from the coupling with the $^{9}$Li spin. 

Finally, exploratory calculations performed for the data at $E=11.1$~MeV/u with extended $^{10}$Li models including $\ell=2$ continuum clearly indicate that these data are not consistent with the presence of a significant $d_{5/2}$ strength below $E_x<4$~MeV. In particular, if $d$-wave resonances are present within this energy range, they are likely to consist of multichannel resonances rather than single-particle states.

\section*{Acknowledgements}
We are grateful to Manuela Cavallaro and Karsten Riisager for useful discussions on the experimental data and their interpretation. This project has received funding from the Spanish Government under project No.~FIS2014-53448-C2-1-P and FIS2017-88410-P and by the European Unions Horizon 2020 research and innovation program under grant agreement No. 654002.

\section*{References}

\bibliography{newbibfile}

\end{document}